\documentclass[sigconf]{acmart}
\AtBeginDocument{%
  \providecommand\BibTeX{{%
    \normalfont B\kern-0.5em{\scshape i\kern-0.25em b}\kern-0.8em\TeX}}}
    
\copyrightyear{2022}
\acmYear{2022}
\setcopyright{acmlicensed}
\acmConference[CIKM '22]{Proceedings of the 31st
ACM International Conference on Information and Knowledge
Management}{October 17--21, 2022}{Atlanta, GA, USA}
\acmBooktitle{Proceedings of the 31st ACM International Conference on
Information and Knowledge Management (CIKM '22), October 17--21, 2022,
Atlanta, GA, USA}
\acmPrice{15.00}
\acmDOI{10.1145/3511808.3557708}
\acmISBN{978-1-4503-9236-5/22/10}

\usepackage[utf8]{inputenc}
\usepackage{amsmath}
\usepackage{caption}
\usepackage{subcaption}
\usepackage{multirow}
\usepackage[labelformat=simple]{subcaption}

\newcommand{\algorithm}{Community and Gender-Aware Seeding} % full name of the algorithm
\newcommand{\algo}{\textsc{CGaS}} % abbreviation of the algorithm
\newcommand{\swap}{swapping} % abbreviation of the swapping mechanism

\begin{document}
\setlength{\textfloatsep}{4pt}
\setlength{\floatsep}{0pt}

\title{%Option 1: SWAP: Making Seeding Algorithms Gender-aware
%Option2: Swapping the Influential: Gender Community Aware Seeding Algorithm\\
 Targeted Influence with Community and Gender-Aware Seeding\\}
 % Research: Community and Gender-Aware Seeding to Influence Targeted Users \\}
%Research: Community-based Approach to Gender-Constrained Influence Maximization \arwen{Need a more attractive title?}}
%%  The format of the title should be the Type (“Research” or “Resource”) followed by a colon, then followed by the title of the paper. For example, for a Resource paper, your title would be Resource: Title. The papers without their types specified in the title may be desk-rejected without review.

\author{Maciej Styczen}
\affiliation{%
  \institution{Swiss Federal Institute of Technology Lausanne (EPFL)}
  \city{Lausanne}
  \country{Switzerland}}
\email{maciej.styczen@epfl.ch}

\author{Bing-Jyue Chen}
\affiliation{%
  \institution{Institute of Information Science}
  \city{Academia Sinica}
  \country{Taiwan}}
\email{chenbingjyue@iis.sinica.edu.tw}

\author{Ya-Wen Teng}
\affiliation{%
  \institution{Institute of Information Science}
  \city{Academia Sinica}
  \country{Taiwan}}
\email{ywteng@iis.sinica.edu.tw}

\author{Yvonne-Anne Pignolet}
\affiliation{%
  \institution{The Dfinity Foundation}
  \country{Switzerland}}
\email{yvonneanne@dfinity.org}

\author{Lydia Chen}
\affiliation{%
  \institution{Department of Computer Science}
  \city{Delft University of Technology}
  \country{Netherlands}}
\email{Y.Chen-10@tudelft.nl}

\author{De-Nian Yang}
\affiliation{%
  \institution{Institute of Information Science}
  \institution{Research Center for Information Technology Innovation}
  \city{Academia Sinica}
  \country{Taiwan}
}
\email{dnyang@iis.sinica.edu.tw}

\begin{abstract}
When spreading information over social networks, seeding algorithms selecting users to start the dissemination play a crucial role.
 The majority of existing seeding algorithms focus solely on maximizing the total number of reached nodes, overlooking the issue of group fairness, in particular, gender imbalance.  To tackle the challenge of maximizing information spread on certain target groups, e.g., females, we introduce the concept of the \emph{community and gender-aware potential} of users. We first show that the network's community structure is closely related to the gender distribution. Then, we propose an algorithm that leverages the information about community structure and its gender potential to iteratively modify a seed set such that the information spread on the target group meets the target ratio. Finally, we validate the algorithm by performing experiments on synthetic and real-world datasets. Our results show that the proposed seeding algorithm achieves not only the target ratio but also the highest information spread, compared to the state-of-the-art gender-aware seeding algorithm.
\end{abstract}




\maketitle

\vspace{-6mm}
\section{Introduction}
% The area of focus of the internship is social network seeding - identifying influential nodes in a social network. More specifically, given a graph representing a social network, find a subset of nodes of a set size (a.k.a. \emph{seed}) such that the total information spread (number of nodes reached through information diffusion process) is maximal. We are particularly interested in a special case of this problem: finding a seed that maximizes the information spread while also ensuring a given gender distribution in the target group.

%% Logic flow:
%% 1. Influence maximization (IM)
%%   1-1. Definition and applications of traditional IM (put related work here)
%%   1-2. Fairness IM (put related work here)
%%   1-3. Disparity IM 
%%
%% 2. Community and gender
%%   2-1. Previous community-based IM approaches (put related work here) achieve high performance to maximize influence. But they do not consider genders.
%%   2-2. However, the homophily effect suggests people with the same gender tend to connect each other (citation needed)
%%   2-3. We argue the need to investigate the role of genders in communities to solve DIM
%%
%% 3. In this paper, we 
%%   3-1. community analysis
%%   3-2. community-based approach to solve DIM
%%
%% 4. List the contributions: community analysis, community-based approach, and experiments

The influence maximization (IM) problem, which chooses a subset of users in the social network as seeds to maximize the number of influenced users \cite{greedy,celf,ris,chen2020efficient,li2018influence,Nguyen:SIGMOD:2016:stop}, has considerable applications, such as viral marketing, political campaigns, and so on. In \cite{greedy}, it has been proved that IM is NP-hard and a greedy algorithm can generate $(1-\frac{1}{e})$-approximation solutions.
However, traditional IM only focuses on the influence spread and neglects the disparity in underrepresented groups (URGs). For example, when the government wants to disseminate a piece of information about financial aid, a good seed group which maximizes the number of influenced individuals but does not take into account if struggling individuals learn about it, may not reach the population which needs it most. 

Extensive research \cite{gershtein2018balanced,group-fairness-in-im,fairness-in-im,ijcai2020-594,Stoica:WWW:2020:SeedingDiver} has investigated numerous IM problem, which aim to provide fairness and deal with the disparity under various definitions, such as maximizing the influence on URGs while ensuring a minimum spread on the whole network \cite{gershtein2018balanced}, maximizing the minimum spread among all groups \cite{group-fairness-in-im}, balancing the numbers of seeds from different groups \cite{fairness-in-im}, and ensuring the ratios of influenced users in respective groups to be the same \cite{ijcai2020-594,Stoica:WWW:2020:SeedingDiver}. 
Recently, the Disparity Influence Maximization (DIM) problem was proposed \cite{cikm2021} to further focus on promoting URG to reach a target ratio, benefiting real-world applications with diverse needs. However, the Disparity Seeding algorithm \cite{cikm2021} proposed to solve DIM is based on the ranking mechanism, neglecting the submodular property of IM and in turn resulting in a limited influence spread.

On the other hand, community-based approaches \cite{cim,community-based-im2,shang2017cofim,kumar2022csr,qiu2019phg} have shown great strengths in solving IM in terms of both effectiveness and efficiency, since they can narrow down the possible seed candidates. Although the above approaches provide promising solutions to traditional IM, none of them has considered fairness. However, as indicated in previous studies \cite{Stoica:WWW:2020:SeedingDiver,avin2015homophily,stoica2018algorithmic}, human social networks usually exhibit homophily, a tendency to favor interactions with similar individuals. For example, females are more likely to interact with females to form communities. Hence, an essential question is how to exploit such communities to improve seed selection so that a specified target gender ratio can be achieved. In this paper, we use the relationships between communities and genders to design a community-based approach to solve DIM.%\footnote{Despite taking a binary gender definition as an example for simplicity, our approach is not limited to the example and can be applied to various definitions of URGs.} 

We start by analyzing a dataset from \textit{Facebook}, where users have three modes of interactions, i.e., likes, comments, and tags. Based on these multi-type interactions, we apply a variant of the Leiden algorithm \cite{leiden} to discover communities and classify them as \textit{male-dominant, female-dominant, and evenly-distributed communities} depending on their majority gender. Our key findings are (a) the community structure is more crucial than the gender for the homophily phenomenon and (b) evenly-distributed communities are usually more influential than the other types of communities.

%%%% shorten it.

% We first study the interaction patterns and the dominant gender in different types of communities in this dataset. Next, we further analyse the interactions between different communities. 
% Our results demonstrate that 1) people indeed tend to interact with those in the same community regardless the type of communities, 2) the influential gender in a community usually complies  with the majority gender, and 3) evenly-distributed communities are usually more influential than the other types of communities, providing evidence that previous community-based approaches that are unaware of genders face difficulties when solving DIM.

Leveraging the observations from our community analysis, we propose \textit{\algorithm\ (\algo)} to improve various seeding approaches that are unaware of genders for solving DIM. \algo\ introduces the \textit{\swap} mechanism to refine the seed group so that the gender distribution of influenced users satisfies the target ratio specified by DIM. To evaluate the influence of a user on different genders, %to foster the \swap\ mechanism, 
\algo\ proposes a novel measure, the \textit{Gender-aware Potential Influence (GPI)}. %according to homophily and our observations.
Specifically, to gain more influenced users of the target gender, \algo\ replaces the seeds having a small GPI on the target gender with users having a large GPI on the other gender while retaining the total influence spread. Note that \algo\ offers great flexibility since it is applicable to refine the seed groups discovered by various seeding algorithms.
Our experiments on the \textit{Facebook} dataset demonstrate that \algo\ can achieve the specified target gender ratio and outperforms Disparity Seeding~\cite{cikm2021} in terms of the influence spread. Our contributions include:

\begin{itemize}
    \item %To the best of our knowledge, 
    We are the first to study the affinity between communities and genders in disparity influence maximization.%to design a community-gender aware seeding to solve DIM.
    \item We develop a novel community-gender aware seeding algorithm, \algo\, which iteratively refines the seeds based on a new metric, the Gender-aware Potential Influence (GPI). %to measure a user's influence on different genders. 
    %flexible \swap\ mechanism to refine the seed group for various seeding algorithms and
    \item The experimental results show that \algo\ achieves more than two times the influence spread of Disparity Seeding on the Facebook dataset.
\end{itemize}

\vspace{-3mm}
\section{Community Analysis}
\label{sec:analysis}

\subsection{Dataset and community detection}
\label{sec:fb}
% \begin{table}[t]
%     \caption{Facebook gender distribution: senders and receivers.}
%     \label{tab:gender}
%     \centering\small
%     \begin{tabular}{c cc c cc}
%     \hline
%     {Interaction} & \multicolumn{2}{c}{Sender} && \multicolumn{2}{c}{Receiver}\\ \cline{2-3}\cline{5-6}
%     & Male & Female && Male & Female \\ 
%     \hline \hline
% Like & 41.1\% (750) & 58.9\% (1079) && 40.9\% (743) & 59.1\% (1074) \\ 
% Comment & 40.9\% (750) & 59.1\% (1087) && 40.7\% (735) & 59.3\% (1073) \\ 
% Tag & 40.6\% (704) & 59.4\% (1031) && 37.3\% (540) & 62.7\% (910) \\ 
% All & 40.8\% (763) & 59.2\% (1104) && 40.9\% (748) & 59.1\% (1082) \\ 
%     \hline
%     \end{tabular}
% \end{table}

The data was gathered from voluntary senior students at 25 university departments through the Facebook API \cite{cikm2021}.\footnote{More information about the dataset can be found at \url{https://www2.ios.sinica.edu.tw/sc/en/home2.php}.} Specifically, the dataset contains 1870 unique users (765 males and 1,105 females) and around 20 million interactions, each of which records users' genders and the interaction type (i.e., likes, comments, or tags). %The students' profiles (e.g., academic standing and hometown) are also collected by questionnaires. 
The period of interactions spans from March 2008 to May 2016. %, and more than 97\% of interactions are after August 2012. %Table \ref{tab:gender} shows the main statistics of how males or females interact with each other through three types of interactions. Note that a higher female number than male one of both sender and receiver is observed in this dataset.

Following \cite{multigraph_community}, we first discover communities by leveraging a variant of Leiden algorithm \cite{leiden}, which is suitable for multi-type interaction social networks. %\footnote{Note that we jointly consider three types of interactions for community detection because of the dependence of different types. For instance, if one user likes another one's post, it is likely that the user also comments on their post.}
Next, according to two-third majority known in politics \cite{Robert2000}, we define a community as a \textit{male-dominant (or female-dominant) community} if the number of males (or females) is more than twice that of females (or males) in this community. Otherwise it is an \emph{evenly-distributed community}.
%we define a \emph{gender-dominant community} to be a community with Shannon entropy less than $0.9183$ (i.e., the user number of one gender is more than twice that of the other gender); otherwise it is called a \emph{evenly-distributed community}. For instance, if there are half of female users in a community, the community is evenly-distributed because its entropy is $-\frac{1}{2}\log_2(\frac{1}{2})-(1-\frac{1}{2})\log_2(1-\frac{1}{2})=1>0.9183$. 
As a result, we obtain $44$ communities with an average size of $42.2$ ($\pm23.49$) members, includiung $21$ evenly-distributed (Even), $16$ female-dominant (Female), and $7$ male-dominant communities (Male).

\begin{figure}[t]
    \begin{center}
    \begin{subfigure}[b]{0.23\textwidth}
        \includegraphics[width=\textwidth]{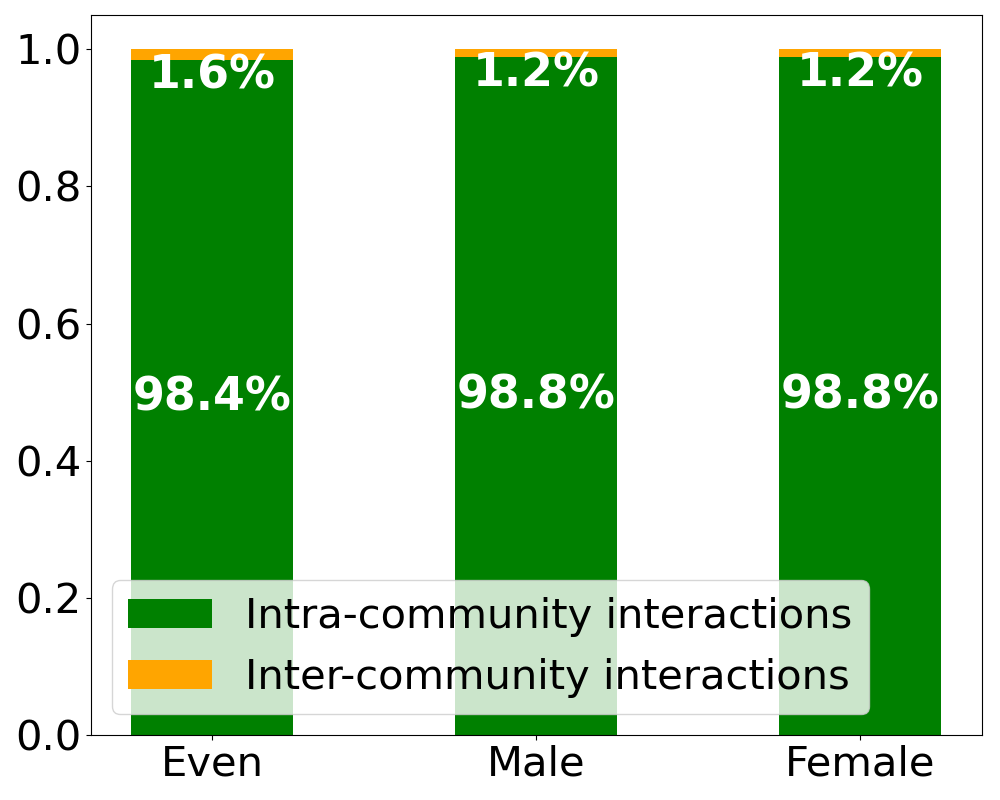}
        \caption{The distribution of inter-/intra-community interactions.}
        \label{fig:user_interaction}
    \end{subfigure}\hfill
     \begin{subfigure}[b]{0.23\textwidth}
         \centering
         \includegraphics[width=\textwidth]{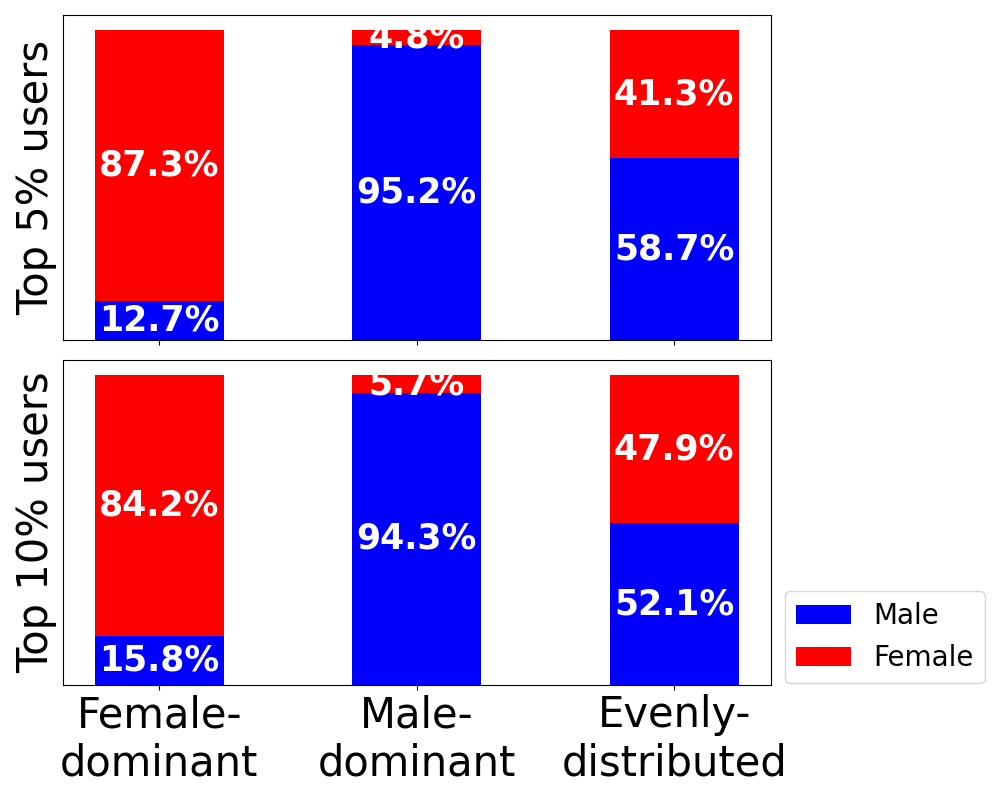}
         \caption{The gender distribution of the top-ranked users.}
         \label{fig:gender_distribution}
     \end{subfigure}\hfill
    \vspace{-3mm}
    \caption{Distributions of user interactions for tagging.}%, evaluated for communities with an even, male or female-dominated population.}%\lc{Add the explanation of inter and intra in a)}}
    \label{fig:interaction}
    \end{center}
\end{figure}

\begin{figure}[t]
    \begin{center}
    \begin{subfigure}[b]{0.23\textwidth}
         \centering
         \includegraphics[width=\textwidth]{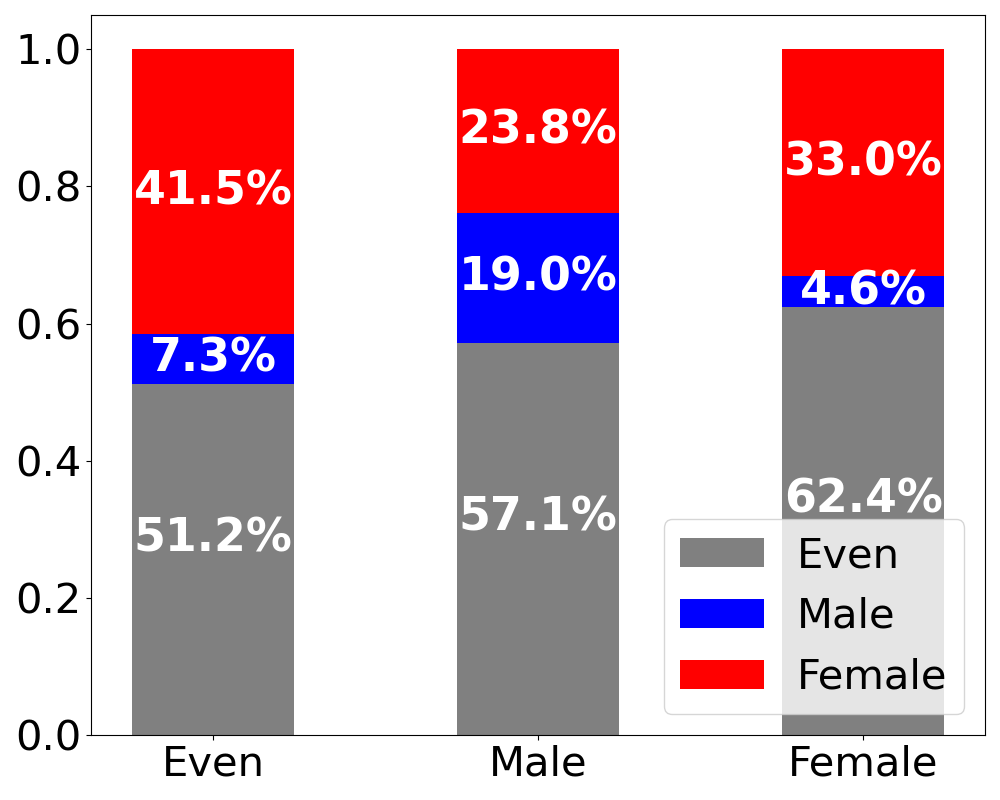}
         \caption{Community interaction.}
         \label{fig:comm_interaction}
     \end{subfigure}\hfill
    \begin{subfigure}[b]{0.23\textwidth}
        \includegraphics[width=\textwidth]{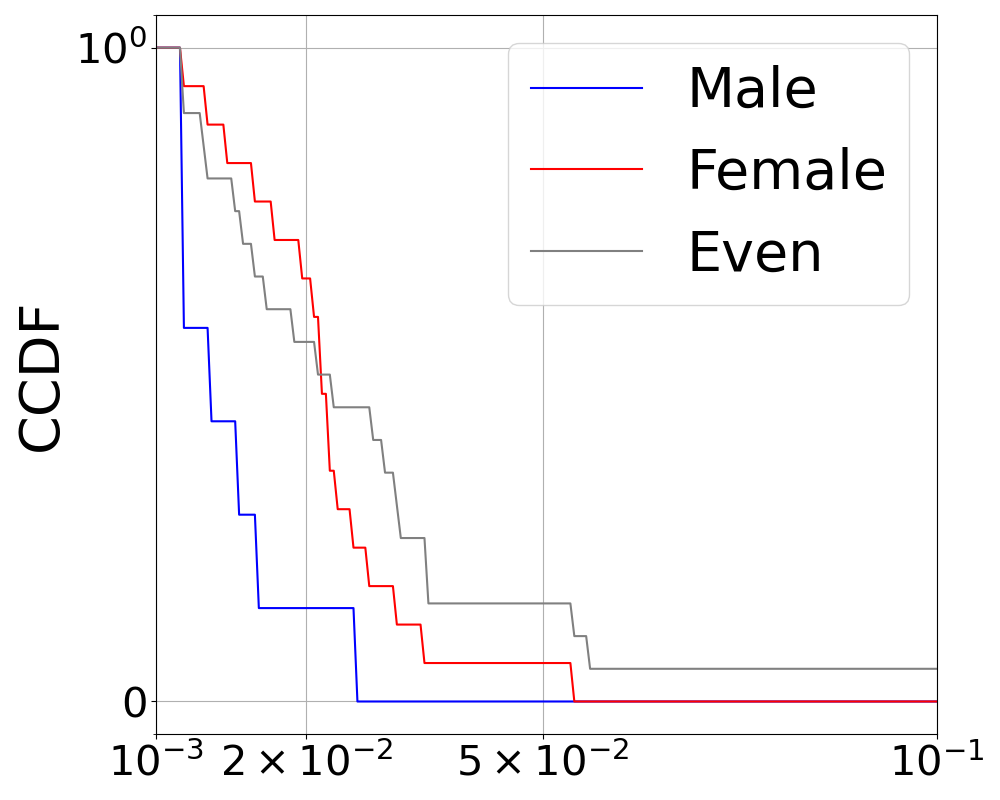}
        \caption{The CCDF of PageRank.}
        \label{fig:page_ccdf}
    \end{subfigure}\hfill
     \vspace{-3mm}
    \caption{Inter-community interactions for tagging.}
    \label{fig:pagerank}
    \end{center}
\end{figure}

\subsection{User interaction}
\label{sec:user}

We first examine the interaction pattern of users in different types of communities in terms of the type of communities their objects belong to. Due to the space limitation, we follow \cite{cikm2021} to focus on the analysis of tags. Figure~\ref{fig:user_interaction} shows the percentage of different types of communities that users in one specific type of community interact with. Take the second bar (i.e., Male) as an example, which shows that the users in male-dominant communities receive and send most tags to users in the same community. Generally, the results demonstrate that more than $98\%$ of users tend to interact with users in the same community regardless the type %or direction of interactions and the gender distribution 
of communities. %Note that the distributions of likes and comments are almost the same as that of tags.

We then analyze whether the gender distribution of top-ranked users (i.e., those who interact with intra-community users the most) in a community complies with the gender that dominates the community. We examine the top $5\%$ and top $10\%$ users as the top-ranked users in Figure~\ref{fig:gender_distribution}. The results show that the gender distribution of the top-ranked users generally follows the dominant gender of the community. For instance, there are on average $84.2\%$ female users in the top $10\%$ users of a female-dominant community. %Note that we also observe one exception: there exists a female-dominant community in which about half of the top $10\%$ users are males. 
%This means if we plan to influence a gender-dominant community to improve fairness, it is not necessary to choose users by the gender who dominates the community. Instead, we may choose influential users from the top-ranked users in each community.
The observations suggest that with respect to influence in a gender-dominant community, users of the dominant gender are more likely to influence more members of the community.

% \begin{table}[t]
% \caption{The female ratio in the top-ranked (top 5\% | 10\%) users.}
% \label{tab:top_gender}
% \small
% \begin{center}
% \begin{tabular}{||c c c c||} 
%  \hline
%  Dominance & Like & Comment & Tag \\ 
%  \hline\hline
%  Female & 85.3\% | 84.6\% & 81.4\% | 80.2\% & 87.3\% | 84.2\% \\ 
%  \hline
%  Male & 4.8\% | 14.9\% & 16.7\% | 9.8\% & 4.8\% | 5.7\% \\
%  \hline
%  Even & 50.0\% | 51.2\% & 50.8\% | 49.6\% & 41.3\% | 47.9\% \\
%  \hline
% \end{tabular}
% \end{center}
% \end{table}

\subsection{Inter-community interaction}
\label{sec:inter}
Next, we focus on inter-community interactions among different types of communities. Specifically, we first transform the original user-level social network into a community-level network, where nodes are communities and weighted-edges represent the numbers of interactions between members of two different communities. Figure~\ref{fig:comm_interaction} shows the distribution of interactions among communities, where the number of interactions (i.e., weights) on each edge is considered. %For ease of reference, the rightmost column (Natural) presents the distribution of the three types of communities on this dataset. %of all. 
We find that for all types of communities, users have the least interactions with male-dominant communities. The ratios of users in female-dominant and evenly-distributed communities interacting with those in male-dominant communities (i.e., blue parts) are even lower than that of the male-dominant community on this dataset (i.e., $\frac{7}{44}=15.9\%$), while more interactions are observed in interactions between male-dominant communities. 
By contrast, female-dominant and evenly-distributed communities usually tend to interact with each other, i.e., female-dominant communities have the highest ratio to interact with evenly-distributed communities, and vice versa. According to the above observations, different types of communities indeed have diverse patterns in the type of communities they interact with.

Finally, we report the PageRank \cite{pagerank} scores of each community in the community-level graph to examine which type of communities is more influential. Figure~\ref{fig:comm_interaction} reveals that evenly-distributed communities are typically the most influential since the complementary cumulative distribution function (CCDF) tail (i.e., the top-ranked communities) of \emph{Even} is greater than those of the other twos. Moreover, as shown in Figure~\ref{fig:comm_interaction}, evenly-distributed and female-dominant communities interact with each other very frequently. These observations provide evidence that traditional community-based IM approaches that target on the most influential communities may neglect the influence on the underrepresented group.

%\begin{table}[t]
%\caption{The gender distribution of the top-ranked %users in different types of communities.}
%\vspace{-3mm}
%\label{tab:top_gender}
%\small
%\begin{center}
%\begin{tabular}{||c | c c | c c ||} 
% \hline
% \multirow{2}{*}{Community} & \multicolumn{2}{|c|}%{Top 5\%} & \multicolumn{2}{|c||}{Top 10\%} \\
% & Male & Female & Male & Female \\
% \hline
% Female-dominant & 12.7\% & \textbf{87.3\%} & %15.8\% & \textbf{84.2\%} \\ 
% Male-dominant & \textbf{95.2\%} & 4.8\% & %\textbf{94.3\%} & 5.7\% \\
% Evenly-distributed & 58.7\% & 41.3\% & 52.1\% & %47.9\% \\
% \hline
%\end{tabular}
%\end{center}
%\end{table}

\section{The DIM Problem and Algorithm}
\subsection{Problem formulation}
In this paper, we aim to solve the Disparity Influence Maximization (DIM) problem, which is formally defined as follows. 

\begin{definition}[DIM \cite{cikm2021}]
\label{def:DIM}
Given a social network $G=(V,E)$, a diffusion model, a parameter $k$, a target gender ratio $\zeta$, and an error margin $e$, DIM finds a seed group $S \subseteq V$ with $\left| S \right|=k$ to maximize the influence spread with the constraint that the ratio of the target gender in the influenced users is $\zeta$ within an error margin $e$.
\end{definition}

\subsection{Algorithm}
%\lc{Maciej, add a few sentences about how the algorithms are inspired by the related work}
%% Logic flow
%% 1. Overview (done)
%%   * If we have space and you have time, put a pseudo-code here
%% 2. Gender-aware Potential Influence (done)
%% 3. Swapping mechanism
%%   3-1. Indicate that we do swapping by iteration
%%   3-2. For each iteration, what should be done
%%   3-3. The stopping criteria (make sure the algorithm stops)
%%   * If possible, try to emphasize 
%%     a) we are able to retain a large influence spread as the original algorithm does, and why we can do that, e.g., using users with a larger GPI on g and also a large PI as candidates?
%%     b) the connections with the observations from Section 2.

%% 1. Overview
Leveraging the observations from Section~\ref{sec:analysis}, we propose \textit{\algorithm\ (\algo)} to solve DIM. As seeding algorithms that are unaware of genders fail to achieve the specified influenced ratio of the target gender $\zeta$, \algo\ proposes a \swap\ mechanism to refine the seed group by evaluating users' influence on different genders. To exploit to homophily, \algo\ introduces the \textit{Gender-aware Potential Influence (GPI)} by summing up the intra- and inter-community influence on some gender that the neighbors of this gender can make. \algo\ exchanges seeds and non-seeds according to their GPI in iterations aiming get closer to $\zeta$ while ensuring a large influence spread. In the following, we first formally define GPI and then explain the \swap\ mechanism in detail.

%% 2. Gender-aware Potential Influence
To evaluate one's influence on different genders, \algo\ defines the Gender-aware Potential Influence (GPI) measure. A user $v$'s GPI on a gender $g \in \{\text{M}, \text{F}\}$ is the sum of the influence on $g$ that $v$'s neighbors of $g$ have. Inspired by the success of community-based algorithms \cite{cim,community-based-im2,shang2017cofim,kumar2022csr,qiu2019phg}, GPI considers the \textit{Gender-aware Community Influence (GCI)} to evaluate neighbors' influence on different genders from the intra- and inter-community aspects. Formally, we define the GPI of a user $v$ on a gender $g$ as follows.
\begin{align}
GPI_g(v) = \sum\limits_{w \in N(v) \setminus A, w\text{'s gender is }g} b_{vw} \times GCI_g(w),
\end{align}
where $GCI_g(w)$ is $w$'s GCI on $g$ (defined later), $N(v)$ is the set of $v$'s neighbors, $A$ is the set of influenced users, and $b_{vw}$ is the weight on the edge between $v$ and $w$ (given by the social network).

Analogously to \cite{qiu2019phg}, we evaluate GCI by considering two types of users in communities. %\footnote{\algo\ is flexible to employ various community detection algorithms, e.g., a variant of Leiden \cite{leiden}.} 
One type represents \textit{core users} who only have intra-community interactions, and the other type comprises \textit{boundary users} who interact with users inside and outside the community. For a core user, the GCI on a gender $g$ is based on i) the number of users of $g$ in their community. Moreover, we also add ii) the number of their neighbors of $g$ to account for homophily. For a boundary user, GCI on $g$ is determined by the influence on each community he/she interacts with. Similar to core users, GCI also considers i) the average number of users of $g$ in the communities that a boundary user interacts with and ii) the number of his/her neighbors of $g$. As observed in Section~\ref{sec:inter}, users in different types of communities have diverse patterns in the type of communities they interact with, GCI further takes iii) the number of $g$-dominant communities he/she interacts with into account. Therefore, the GCI of a user $v$ on a gender $g$ is defined as follows.
\begin{align}
GCI_g(v) =
\begin{cases}
U_{g}(v) + D_g(v) & \text{if }v\text{ is a core user} \\
\alpha \cdot AU_{g}(v) + D_g(v) + C_{g}(v) & \text{otherwise}
\end{cases}
,
\end{align}
where $U_g(v)$ is the number of users of $g$ in $v$'s community, $D_g(v)$ is the number of $v$'s neighbors of $g$, $AU_g(v)$ is the average number of users of $g$ in the communities $v$ interacts with, and $C_g(v)$ is the weighted number of communities $v$ interacts with and the weight of each community is the fraction of users of $g$. %Note that we consider the fraction of users of $g$ in a community for $C_g(v)$ instead of a binary value to represent whether it is a $g$-dominate community, leading to a more fine-granular metric. 
Besides, as users tend to interact more with intra-community users (observed in Section~\ref{sec:user}), $AU_g(v)$ may overestimate $v$'s influence and is adjusted by a parameter $0< \alpha \leq 1$ reflecting the ratio of inter-community interactions to all interactions.

%% 3. Swapping mechanism
Equipped with GPI, \algo\ applies the \swap\ mechanism to refine the seed group by iteratively exchanging seeds and non-seeds. Let $S_0$ denote the initial seed group generated by an arbitrary, possibly non-gender-aware, seeding algorithm. For each iteration~$i \geq 1$, \algo\ estimates the influence spread of $S_{i-1}$ on each gender to derive the influenced ratio of the target gender $r_{i-1}$. Based on the comparison between $r_{i-1}$ and $\zeta$ (defined in Definition~\ref{def:DIM}), \algo\ determines the gender that should be mainly influenced in this iteration $i$, denoted as $g_i$. Then, \algo\ chooses $n$ seeds from $S_{i-1}$ with the lowest $GPI_{g_i}$ as \textit{out-candidates} and $n$ nodes from $V \setminus S_{i-1}$ with the highest $GPI_{g_i}$ as \textit{in-candidates} to form $n^2$ exchange pairs, where $n$ is a parameter to control the number of exchange pairs.\footnote{A larger $n$ leads to more exchange pairs, which require more search time but provide a higher chance to reach $\zeta$ at once.} Specifically, \algo\ forms an exchange pair by one in-candidate and one out-candidate and estimates the influence spread on each gender considering $S_{i-1}$ is updated by this exchange pair, i.e., removing the out-candidate from the seed group and adding the in-candidate into the seed group. Consequently, \algo\ selects the exchange pair with the largest influence spread on $g_i$ to update $S_{i-1}$ as $S_i$. The \swap\ mechanism continues until $r_i$ reaches $\zeta$ within the error margin $e$ or the maximum iteration $i_\text{max}$ is reached, i.e., $i = i_\text{max}$.

\vspace{-5mm}
\section{Experiments}
%% Logic flow
%% 1. Setup
%%   1-1. Datasets: Facebook and synthetic
%%   1-2. Baselines:
%%     1-2-1. Not gender-aware: agnostic (intensity), CELF (or RIS?) -> for these algorithms, we equip them with our algorithm so that they can be compared with gender-aware algorithms
%%     1-2-2. Gender-aware: diversity seeding (pagerank), disparity seeding (target hi-index)
%%   1-3. Evaluation metrics and parameters (e.g., seed size, error margin), and other notes (e.g., definition of female ratios, times of Monte-Carlo sampling)
%%
%% 2. Results
%%   * Since the results on different datasets seem to have similar observations, I think we may not separate them as two subsections.
%%   2-1. Ratio and spread: 
%%     2-1-1. We have the least error for all ratios on all datasets since ... (We are better because we are limited to the top ranked males and female?)
%%     * leave footnote somewhere to say extreme ratios are not easy and usually impractical (?)
%%     2-1-2. We also have the largest influence spread since ...(how we can achieve it) -> at least how much times of influence spread over disparity seeding?
%%   2-2. Efficiency in terms of the number of iterations. We find users with high GPI to adjust the influenced ratio. Therefore, we can efficiently find right ones instead of exhaustedly trying every possible seed combinations.

\begin{table}[t]
    \caption{The comparison of influence spread and its female ratios under two community-gender aware seeding.}
    \vspace{-3mm}
    \label{tab:results}
    \setlength{\tabcolsep}{0.13em}
    \small
    \centering
    \begin{tabular}{||c c | c c c | c c c | c c | c c ||} 
     \hline
     \multicolumn{2}{||c|}{Specified} & \multicolumn{3}{|c|}{AN+\algo} & \multicolumn{3}{|c|}{CELF+\algo} & \multicolumn{2}{|c|}{DV} &  \multicolumn{2}{|c||}{DP} \\
     \multicolumn{2}{||c|}{ratio ($\zeta$)} & Ratio & Spread & \#Iter & Ratio & Spread & \#Iter & Ratio & Spread & Ratio  & Spread \\
     \hline
     \multirow{4}{*}{\rotatebox[origin=c]{90}{Facebook}} & 0.4 & \underline{.408} & \textbf{245.2} & 19 & .461 & 337.0 & 20 & .621 & 223.7 & \underline{.403} & 153.6\\ 
     & 0.5 & \underline{.500} & 228.2 & 6 & \underline{.508} & \textbf{381.8} & 4 & .621 & 223.7 & \underline{.506} & 112.4 \\
     & 0.6 & \underline{.607} & 221.7 & 3 & \underline{.594} & \textbf{413.3} & 2 & .647 & 222.4 & \underline{.595} & 191.9 \\
     & 0.7 & \underline{.699} & 251.9 & 8 & \underline{.692} & \textbf{381.2} & 12 & .663 & 219.2 & \underline{.698} & 165.2 \\
     \hline
     \multirow{4}{*}{\rotatebox[origin=c]{90}{synthetic}} & 0.3 & \underline{.305} & \textbf{41.48} & 5 & \underline{.295} & 37.50 & 5 & .380 & 44.07 & \underline{.305} & 26.35 \\ 
     & 0.4 & \underline{.401} & \textbf{41.63} & 2 & \underline{.396} & 39.51 & 2 & .412 & 45.91 & \underline{.391} & 36.67  \\
     & 0.5 & \underline{.500} & \textbf{37.01} & 5 & \underline{.502} & 36.25 & 5 & .448 & 44.68 & .548 & 33.12  \\
     & 0.6 & \underline{.603} & 30.77 & 7 & \underline{.604} & \textbf{32.77} & 6 & .481 & 42.73 & \underline{.592} & 27.08  \\
     \hline
    \end{tabular}
\end{table}

\subsection{Setup}
\noindent\textbf{Datasets.}
We first use the tag interactions on the Facebook dataset (Section~\ref{sec:fb}) to evaluate the performance of \algo. Furthermore, we construct the following synthetic dataset to illustrate our approach on a smaller network. As there are more females than males in the Facebook dataset (i.e., the female ratio is 60\%), we apply a modified stochastic block model \cite{holland1983stochastic} to generate a synthetic dataset with a community structure that contains more males than females. %Following \cite{} \arwen{Add references to justify these parameters}, 
This dataset consists of four communities with 28, 20, 22, and 30 users, and the corresponding female ratios are 0.8, 0.5, 0.25, and 0.75, respectively. In other words, the synthetic dataset has 60 males and 40 females and the sizes of the communities are in the same range as small communities in the Facebook dataset. To mimic the interaction patterns of social networks like the Facebook dataset, the probabilities of having intra- and inter-community interactions are chosen in ranges $[0.7, 0.8]$ and $[0.01, 0.03]$, respectively. 

\noindent\textbf{Baselines.}
We use two non-gender-aware seeding algorithms and two gender-aware seeding algorithms for comparison. i) Agnostic seeding (AN) \cite{Stoica:WWW:2020:SeedingDiver} selects the top-ranked users as seeds according to the number of users' interactions. ii) CELF \cite{celf} greedily selects the users with the maximum marginal gain on the influence spread as seeds. iii) Diversity seeding (DV) \cite{Stoica:WWW:2020:SeedingDiver} selects the top-ranked males and females as seeds according to the number of users' interactions, where the female ratio of the seed group is searched between the specified female ratio $\zeta$ and the female ratio of AN's seed group.
iv) Disparity seeding (DP) \cite{cikm2021} selects the top-ranked males and females as seeds according to the Target HI-Index scores, where the female ratio of the seed group is learned from a scaling function. As AN and CELF are not aware of genders, we exploit \algo\ to refine the seed groups derived from them, denoted as AN+\algo\ and CELF+\algo, to demonstrate the applicability of \algo\ to different seeding algorithms. Moreover, we compare AN+\algo\ and CELF+\algo\ with DV and DP to show the superiority of \algo\ to maximize the influence spread while approaching $\zeta$.

\noindent\textbf{Metrics and Parameter Settings.}
We evaluate i) the female ratio of the influenced users and ii) the influence spread by varying the specified female ratios of the influenced users $\zeta$. The sizes $k$ of the seed group of the Facebook and synthetic datasets are set to 25 and 10, respectively, and the error margin for both datasets is set to $0.01$. Following~\cite{Stoica:WWW:2020:SeedingDiver}, the diffusion is simulated through the IC model~\cite{greedy} over 10000 times, where the probability of user~$v$ influencing user~$w$ (i.e., the weight $b_{vw}$) is the number of interactions $w$ receives from $v$ divided by the total number of interactions $w$ receives \cite{Nguyen:SIGMOD:2016:stop}. For \algo, $\alpha$, $n$ and $i_\text{max}$ are set as $\frac{1}{3}$, 5 and 20 for both datasets.

\vspace{-2mm}
\subsection{Results}
Table~\ref{tab:results} lists the female ratio of the influenced users and the influence spread achieved by different algorithms on the Facebook and synthetic datasets. The female ratios with underlines represent the success of achieving $\zeta$ within the error margin $e$, while the spreads with bold are the maximum spread among the algorithms having female ratios with underlines. Note that we vary $\zeta$ between $[0.4,0.7]$ and $[0.3, 0.6]$ for the Facebook and synthetic datasets, respectively, since their dominant genders are different.%\footnote{We do not examine extreme ratios since the goal of real-world applications is still to maximize the influence and an extreme ratio usually prevents achieving this goal. Hence, setting an extreme ratio is somewhat impractical.}
From the results, we have three observations. (1) \algo\ can approximate $\zeta$ based on the seed groups obtained by AN and CELF, showing the applicability of \algo\ to various seeding algorithms. (2) DV fails to achieve $\zeta$ within $e$ and has moderate errors only when $\zeta$ is close to the female ratio in the population (i.e., 0.6 and 0.4 in the Facebook and synthetic datasets, respectively). By contrast, AN+\algo, CELF+\algo, and DP deviate least from $\zeta$. (3) Although DP has small errors, AN+\algo\ and CELF+\algo\ always exhibit a much larger influence spread than DP while ensuring $\zeta$ is reached, since \algo\ exploits the intra- and inter-community influence to evaluate the users' influence. Specifically, CELF+\algo\ achieves more than two times the influence spread of DP on the Facebook dataset.

Table~\ref{tab:results} further presents the number of iterations required by \algo\ to approach $\zeta$. As $\zeta$ deviates more from the gender distribution of the population, %(i.e., smaller on the Facebook dataset and larger on the synthetic dataset), 
more iterations are required. Since non-gender-aware seeding algorithms usually yield the gender distribution in the influenced users following that in the population, a larger difference between $\zeta$ and the gender distribution of the population results in more efforts to adjust the seed group. For example, CELF+\algo\ is unable to reach $0.4 \pm 0.01$ within $i_\text{max}=20$.

\section{Conclusion}
Motivated by the prevalent community structure and the asymmetric influence of different genders in social networks, we conducted a first-of-its-kind gender-aware community analysis and presented a novel seeding algorithm to promote the information spread on the target group. We studied how the community structure is affected by the gender distribution and showed that many communities have a strong affinity to a single gender. Secondly, we designed a community-gender aware algorithm that achieves a target ratio of information spread by iteratively adjusting the seeding selection based on the community-gender structure. We evaluated the proposed seeding algorithm against the state-of-the-art gender-aware seeding algorithms on both synthetic graphs and a Facebook trace. The results demonstrate that the proposed community-gender aware seeding algorithm achieves the target influence ratio while maximizing the information spread.

\section*{Acknowledgement}
This work is supported in part by MOST under grants 110-2221-E-001-014-MY3, and 109-2221-E-001-017-MY2. We thank NCHC of NARLabs in Taiwan for providing computational and storage resources.

\clearpage

\bibliographystyle{ACM-Reference-Format}
\bibliography{ref}

\end{document}